\documentclass[10pt]{JHEP3}

\usepackage{amsmath}

\newcommand{\e}{\epsilon}

\renewcommand{\L}{{\mathcal{L}}}
\newcommand{\bL}{\bar{{\mathcal{L}}}}
\newcommand{\z}{{\bar z}}

\renewcommand\O{{\mathcal{O}}}
\newcommand{\be}[1]{ \begin{equation}\label{#1} }
\newcommand{\ee}{\end{equation}}
\newcommand{\ben}[1]{\begin{eqnarray}\label{#1} }
\newcommand{\een}{\end{eqnarray}}
\newcommand{\eq}[1]{(\ref{#1})}
\newcommand{\p}{\partial}

\newcommand{\refb}[1]{(\ref{#1})}

\title{Topologically Massive Gravity and Galilean Conformal Algebra: A Study of Correlation Functions}

\author{
Arjun Bagchi\\
$\;$ $\,$School of Mathematics, \\
$\;$ $\,$University of Edinburgh \\
$\;$ $\,$Kings Buildings, Edinburgh EH9 3JZ\\
$\;$ $\,$United Kingdom\\

$\;$\email{arjun.bagchi@ed.ac.uk}
}

\abstract{The Galilean Conformal Algebra (GCA) arises from the conformal algebra in the non-relativistic limit. 
In two dimensions, one can view it as a limit of linear combinations of the two copies Virasoro algebra. 
Recently, it has been argued that Topologically Massive Gravity (TMG) realizes the quantum 2d GCA in a particular 
scaling limit of the gravitational Chern-Simons term. To add strength to this claim, we demonstrate a matching of 
correlation functions on both sides of this correspondence. A priori looking for spatially dependent 
correlators seems to force us to deal with high spin operators in the bulk. We get around this difficulty by constructing 
the non-relativistic Energy-Momentum tensor and considering its correlation functions. On the gravity side, our
analysis makes heavy use of recent results of Holographic Renormalization in Topologically Massive Gravity.}

\preprint{EMPG-10-27}

\begin{document}

\baselineskip 3.5ex

\section{Introduction}

The AdS/CFT conjecture \cite{Maldacena:1997re} has been the cornerstone in the research in string theory over the past decade. Recently, there has been a
flurry of activity in the community in applying the correspondence to potential real life condensed matter systems. Relatedly, 
there has been a growing interest in non-relativistic versions of AdS/CFT. The most studied of these versions has been the ones with 
Schrodinger symmetry algebra, the largest symmetry algebra of the free Schrodinger equations \cite{Hagen:1972pd, Niederer:1972zz, Henkel:1993sg}. 
This symmetry is known to be realized in cold atoms at unitarity \cite{Nishida:2007pj}. A gravity dual for these systems was proposed in \cite{Son:2008ye, Balasubramanian:2008dm}, 
following which quite a body of literature has built up. We refer the reader to the excellent review \cite{Hartnoll} 
and the references therein for a flavour of the activity in this direction. 

In this paper, we would focus on the version of non-relativistic AdS/CFT which can be best motivated as the true non-relativistic limit
of the conjecture \cite{Bagchi:2009my, Bagchi:2009ca, Bagchi:2009pe}. The relativistic conformal algebra on the boundary of AdS is systematically reduced to what we call the Galilean 
Conformal Algebra by a process of parametric group contraction. This algebra is surprising in many aspects, the most intriguing of 
which is the fact that it can be given an infinite lift in any space-time dimensions. The GCA is also important to non-relativistic 
hydrodynamics. The finite algebra turns out to be the symmetry algebra of the incompressible Euler equations and a part of the full 
infinite algebra is also realized as its symmetries. 

Given the surprising infinite dimensional lift of the GCA for all spacetime dimensions, it is natural to first figure out what the
story is in two dimensions. As is well known, the case of D=2 is special because here the relativistic conformal algebra gets enhanced 
to two copies of the infinite dimensional Virasoro algebra. One would expect the infinite GCA to be related to these two copies of the
relativistic Virasoro algebra and this expectation was borne out by our analysis in \cite{Bagchi:2009pe}. The GCA emerges by taking 
simple linear combinations of the two copies of the Virasoro algebra and then looking at the non-relativistic limit. 
One can also look to the quantum aspects of the GCA in two dimensions in the same spirit as Virasoro algebra. The central charges for 
the GCA are asymmetric, they are linear combinations of the parent relativistic central charges. 

The initial bulk description of the GCA was given in terms of a novel Newton-Cartan like $AdS_2 \times R^d$ in \cite{Bagchi:2009my}.
The bulk metric degenerates in this limit and one can formulate the gravity theory in terms of dynamical Christoffel symbols in a 
geometrized version of Newtonian gravity. Unlike in the usual Newton-Cartan formulation of flat space where one had a specially selected
time direction, in AdS both the radial and temporal directions survive the scaling giving a fibre bundle structure where the base is 
an $AdS_2$ made out of the surviving radial and time directions and there are flat spatial fibres over this base. 

For the bulk theory dual to the two-dimensional boundary GCA, one would need to consider such a Newton-Cartan like $AdS_2 \times R$ 
emerging from an $AdS_3${\footnote {Recently, in \cite{Bagchi:2010xw}, we constructed metrics which realized the global 
2d GCA as the isometry-algebra. For the algebra in two spacetime dimensions, the metrics in four and five dimensions 
turned out to have exotic signatures. This further suggests that the Newton-Cartan structure is the best framework for 
describing the non-relativistic bulk.}}. The natural candidate for trying to model the asymmetric central charges would be Cosmological Topologically
Massive gravity. This contains a vacuum $AdS_3$ solution and also an excited state of a BTZ black hole. This has been recently looked at 
in \cite{Hotta}. The authors found that the GCA emerged as the asymptotic symmetry of this structure in the non-relativistic limit. 
The asymmetrical central charges for the GCA was realized by making the coefficient of the gravitational Chern-Simons term very large.
The authors also found a non-relativistic generalization of Cardy's entropy formula for the BTZ blackhole in by considering this limit. 

In this paper, we find further evidence that this connection is true. We look at correlation functions on both sides of the duality. 
At first sight this is a complicated problem. The construction of the correlation functions of primary operators of the boundary GCA
in \cite{Bagchi:2009ca} emphasized that if we are to look for non-trivial spatial dependence, then we would need to focus on the high 
spin sector. In the dual gravity side, computing correlators for arbitrarily high spins is at best cumbersome. We get around this 
apparently insurmountable problem by making an observation about energy-momentum tensors which leads to a gross simplification. 
The calculation on the field theory side follows from the systematic theory of limits established in \cite{Bagchi:2009pe} and the 
corresponding gravity calculation makes heavy use of recent work in holographic renomalization of CTMG in \cite{Skenderis:2009nt}.

The rest of the paper is organized as follows. In Sec 2, we revisit the Galilean Conformal Algebra in general dimensions and then 
concentrate on the results in two dimensions. In Sec 3, we focus our attention on the construction of the non-relativistic energy momentum
tensors and describe a few of their properties. In Sec 4, we start discussing the bulk theory. We revisit the simplified
case where usual holography works for correlation function \cite{Alishahiha:2009np} and then emphasize why we need to
look at Topologically Massive Gravity, briefly describing the construction of \cite{Hotta}. In Sec~5, after a 
brief summary of holographic renormalization, the results of \cite{Skenderis:2009nt} are discussed. 
We then apply these results to reproduce the calculations in the boundary theory and add support to the 
claim of \cite{Hotta}.

\section{A Review of the GCA}

\subsection{GCA in arbitrary dimensions}

The maximal set of conformal isometries of Galilean spacetime generates the infinite dimensional Galilean Conformal
Algebra \cite{Bagchi:2009my}. The notion of Galilean spacetime is a little subtle since the spacetime metric degenerates 
into a spatial part and a temporal piece. Nevertheless there is a definite limiting sense (of the relativistic spacetime) 
in which one can define the conformal isometries (see \cite{Duval:2009vt}) of the nonrelativistic geometry. Algebraically, 
the set of vector fields generating these symmetries are given by
\ben{gcavec}
L^{(n)} &=& -(n+1)t^nx_i\p_i -t^{n+1}\p_t \,,\cr
M_i^{(n)} &=& t^{n+1}\p_i\,, \cr
J_a^{(n)} \equiv J_{ij}^{(n)} &= & -t^n(x_i\p_j-x_j\p_i)\,,
\een 
for integer values of $n$. Here $i=1\ldots (d-1)$ range over the spatial directions. 
These vector fields obey the algebra
\ben{vkmalg}
[L^{(m)}, L^{(n)}] &=& (m-n)L^{(m+n)}, \qquad [L^{(m)}, J_{a}^{(n)}] = -n J_{a}^{(m+n)}, \cr
[J_a^{(n)}, J_b^{(m)}]&=& f_{abc}J_c^{(n+m)}, \qquad  [L^{(m)}, M_i^{(n)}] =(m-n)M_i^{(m+n)}. 
\een
There is a  finite dimensional subalgebra  of the GCA (also sometimes
referred to as the GCA) which consists of taking $n=0,\pm1$ for the
$L^{(n)}, M_i^{(n)}$ together with $J_a^{(0)}$. This algebra is obtained 
by considering the nonrelativistic  contraction of the usual (finite dimensional)
 global conformal algebra $SO(d,2)$ (in $d>2$ spacetime dimensions).

\subsection{GCA in 2d}

In two spacetime dimensions, as is well known, the situation is special. 
The relativistic conformal algebra is infinite dimensional
and consists of two copies of the Virasoro algebra. 
One expects this to be related to the infinite dimensional GCA algebra \cite{Bagchi:2009pe}. Indeed in two dimensions the 
non-trivial generators in  \eq{vkmalg} are the $L_n$ and the $M_n$:
\be{gca2dvec}
L_n = -(n+1)t^n x\p_x -t^{n+1}\p_t\,, \quad M_n = t^{n+1}\p_x\,,
\ee
which obey
\be{vkmalg2d}
[L_m, L_n] = (m-n)L_{m+n}\,, \quad [M_{m}, M_{n}] =0\,, \quad [L_{m}, M_{n}] = (m-n)M_{m+n} 
\ee

These generators  in \eq{gca2dvec} arise precisely from a nonrelativistic contraction of the two copies of the Virasoro algebra.
To see this, let us remember that the non-relativistic contraction consists of taking the scaling 
\be{nrelscal}
t \rightarrow t\,, \qquad   x \rightarrow \epsilon x\,,
\ee
with $\epsilon \rightarrow 0$. This is equivalent to taking the velocities $v \sim \epsilon$ to zero
(in units where $c=1$).Consider the vector fields which generate the centre-less Virasoro Algebra in two dimensions :
\be{repn2dV}
\L_n = -z^{n+1} \p_z\,, \quad \bL_n = -\z^{n+1} \p_{\z}\,.
\ee 
In terms of space and time coordinates, $z= t+x$, $\z=t-x$. Expressing $\L_n , \bL_n$ in terms of $t,x$ and taking the above scaling  
\eq{nrelscal} reveals that in the limit the combinations
\be{GCArepn}
\L_n + \bL_n = -t^{n+1}\p_t - (n+1)t^nx \p_x + \O(\e^2); \quad \L_n - \bL_n = -{1\over \e}t^{n+1} \p_x + \O(\e)\,.
\ee 
Therefore we see that as $\e\rightarrow 0$
\be{Vir2GCA}
\L_n + \bL_n \longrightarrow L_{n}\,, \quad \e (\L_n - \bL_n) \longrightarrow - M_{n}\,.
\ee

At the quantum level the two copies of the Virasoro get respective central extensions
\ben{relalg}
[\L_m, \L_n] &=& (m-n) \L_{m+n} + {c \over 12} m(m^2-1) \delta_{m+n,0}\,, 
\nonumber \\[1mm]
[\bL_m, \bL_n] &=& (m-n) \bL_{m+n} 
+ {\bar c \over 12} m(m^2-1)\delta_{m+n,0}\,.
\een 
Considering the linear combinations \eq{GCArepn} which give rise to the GCA generators as in \eq{Vir2GCA}, we find 
\ben{gcawc}
[L_{m}, L_{n}] &=& (m-n) L_{m+n} + C_1 m(m^2-1) \delta_{m+n,0}\,, \crcr 
[L_{m}, M_{n}] &=& (m-n) M_{m+n} + C_2 m(m^2-1) \delta_{m+n,0}\,, \crcr 
[M_{m}, M_{n}] &=& 0\,.
\een
This is the centrally extended GCA in 2d. Note that the relation between central charges is 
\be{centch}
C_1 = {{c+\bar c} \over 12}\,, \qquad {C_2 \over \e} = {{\bar c-c} \over 12}\,.
\ee
Thus, for a non-zero $C_2$ in the limit $\e\rightarrow 0$ we see that we need $\bar c-c \propto
\O({1\over \e})$. At the same time requiring $C_1$ to be finite we find
that  $c+\bar c$ should be $\O(1)$. Thus (\ref{centch}) can hold only if $c$ and $\bar c$ are large (in the limit $\e\rightarrow 0$)
 and opposite in sign. This immediately implies that the original 2d CFT on which we take the non-relativistic limit cannot be 
unitary. This is, of course, not a problem since there are many statistical mechanical models which are described at a fixed point 
by non-unitary CFTs.

\subsection{Representations of the 2d GCA}

We will construct the representations of the GCA by considering the states having definite scaling dimensions \cite{Bagchi:2009ca, Bagchi:2009pe}:
\begin{equation}
L_0 |\Delta \rangle = \Delta | \Delta \rangle \,.
\label{L0=Delta}
\end{equation}
Using the commutation relations (\ref{gcawc}), 
we obtain 
\begin{equation}
L_0 L_n | \Delta \rangle = (\Delta - n) L_n | \Delta \rangle, \quad 
L_0 M_n | \Delta \rangle = (\Delta - n) M_n | \Delta \rangle. 
\end{equation}
Then the $L_{n}, M_{n}$ with $n >0$ lower the value of the scaling
dimension, while those with $n<0$ raise it. If we demand that the
dimension of the states be bounded from below then we are led to
defining primary states in the theory having the following properties : 
\begin{equation}
L_n|\Delta \rangle =0\,, \quad 
M_n|\Delta \rangle =0\,,
\label{primop} 
\end{equation}
for all $n>0$. 
Since the conditions (\ref{primop})
are compatible with $M_0$ in the sense
\begin{equation}
L_n M_0 |\Delta \rangle = 0\,, \quad 
M_n M_0 |\Delta \rangle = 0\,,
\end{equation}
and also since $L_0$ and $M_0$ commute,
we may  introduce an additional label,
which we will call ``rapidity'' $\xi$:
\begin{equation}
M_0 |\Delta, \xi \rangle = \xi |\Delta, \xi \rangle\,.
\end{equation}

Starting with a primary state $|\Delta,\xi \rangle$, 
one can build up a tower of operators by the action of 
$L_{-n}$ and $M_{-n}$ with $n>0$. The above construction is quite analogous to that of the 
relativistic 2d CFT. In fact, from the viewpoint of the limit (\ref{Vir2GCA})
we see that the two labels $\Delta$ and $\xi$ are related 
to the conformal weights in the 2d CFT as
\be{delxi}
\Delta=\lim_{\epsilon \to 0}
(h+\bar h)\,, \qquad  \xi= \lim_{\e\to 0} \e ({\bar{h} -h})\,,
\ee 
where $h$ and $\bar h$ are the eigenvalues of $\L_0$ and $\bL_0$, respectively. 
We will assume that the operator state correspondence 
in the 2d CFT gives a similar correspondence between the states and the operators
in the GCA:
\be{stateop}
\O (t,x) \leftrightarrow \O(0)|0\rangle\,, 
\ee
where $|0 \rangle$ would be the vacuum state
which is invariant under the generators 
$L_0, L_{\pm1}$, $M_0, M_{\pm1}$. 

\subsection{Two Function of the GCA}

The constraints from the Ward identities for the global transformations $L_{0,\pm 1}, M_{0, \pm 1}$
apply to primary GCA operators.

Therefore consider the two point function of primary operators $\O_1(t_1, x_1)$ and $\O_2(t_2, x_2)$ of conformal and 
rapidity weights $(\Delta_1, \xi_1)$ and $(\Delta_2, \xi_2)$ respectively.
\be{2pt}
G_{\rm GCA}^{(2)}(t_1, x_1, t_2, x_2) 
= \langle \O_1(t_1,x_1) \O_2(t_2,x_2) \rangle \,.
\ee 
The correlation functions only depend on differences of the coordinates $t_{12} = t_1 - t_2$ 
and $x_{12} = x_1 - x_2$ because of the translation symmetries $L_{-1}$ and $M_{-1}$. 
The remaining symmetries give four more differential equations which constrain the answer to be 
\cite{Bagchi:2009ca}
\be{2ptgca}
G_{\rm GCA}^{(2)}(\{t_i, x_i \}) 
= C_{12} \delta_{\Delta_1,\Delta_2} 
\delta_{\xi_1, \xi_2} t_{12}^{-2\Delta_1} 
\exp\left( {2\xi_1 x_{12}\over t_{12}} \right).
\ee 
Here $C_{12}$ is an arbitrary constant, which we can always take to be one by choosing the normalization of the operators.
We can similarly construct the three point function which again is fixed by the symmetries upto an overall constant. 

The GCA two and three point functions can also be obtained by taking an appropriate scaling  limit of the usual 2d CFT answers.
This limit requires scaling the quantum numbers of the operators as (\ref{delxi}), along with the non-relativistic limit
for the coordinates \eq{nrelscal}.

Let us study the scaling limit of the two point correlator.
\ben{R2pt}
G^{(2)}_{\rm 2d\,CFT} 
&=& \delta_{h_1, h_2} \delta_{{\bar h}_1, {\bar h}_2}  
z_{12}^{-2 h_1} \bar z_{12}^{-2 \bar h_1} 
\nonumber \\
&=& 
\delta_{h_1,h_2}\delta_{\bar h_1,\bar h_2}\,\,
t_{12}^{-2h_1} 
\Big( 1 + \epsilon {x_{12} \over t_{12}} \Big)^{-2h_1} 
\,t_{12}^{-2\bar h_1} 
\Big(1 - \epsilon {x_{12} \over t_{12}} \Big)^{-2 {\bar h}_1}
\nonumber \\
&=& \delta_{h_1, h_2} \delta_{{\bar h}_1, {\bar h}_2} \,\ 
t_{12}^{- 2(h_1+\bar h_1)} 
\exp\Big( 
- 2 (h_1- \bar h_1) 
\big(\epsilon { x_{12} \over t_{12}} 
+ {\cal O}(\epsilon^2)\big)\Big)\,.
\een
Now by taking the scaling limit as (\ref{delxi}), 
we obtain the GCA two point function
\be{lim2pt} 
\lim_{\epsilon \to 0} G^{(2)}_{\rm 2d\,CFT} 
=  \delta_{\Delta_1, \Delta_2} 
\delta_{{\xi}_1, {\xi}_2} \, t_{12}^{-2\Delta_1} 
\exp \Big({2 \xi_1 x_{12} \over {t_{12}}} \Big) 
= G^{(2)}_{\rm GCA}\,.
\ee
A similar analysis yields the three point function of the GCA from the relativistic three point function.

\section{The Non-relativistic Energy-Momentum Tensor}

From the above, we saw that we have non-trivial correlation functions (non-trivial in the sense of spatial dependence) 
only when we look at high values of the spin $h - \bar h$ (in order to get a non-zero $\xi$). Our ultimate aim in this paper is to construct a matching 
of non-trivial correlation functions from the boundary and the bulk. It seems naively, that one needs
to consider fields of arbitrary high spin propagating in the bulk if one wants to recover some spatially dependent answers. 
This is a complicated exercise. We also wish to probe the quantum aspects of the GCA. There is the additional drawback 
that the two and three point correlation function of the above described quasi-primary fields would be independent of the central charge.
So it seems that one would need to look at four or higher point functions of the primary operators with arbitrarily high
spin to address the GCA correlation functions in their full generality.   

The dependence on central charges has an obvious answer: looking at the energy-momentum tensor. It turns out that 
if one constructs the non-relativistic EM tensor properly, one gets spatially dependent correlation functions. 
The focal point of the argument is that one must remember that in this non-relativistic limit, we also scale 
the central charges of the relativistic theory (see \refb{centch}). If we focus in the energy-momentum tensor, 
as we would do now, one can easily show that there are spatial dependent terms in the correlation functions, 
even thought we are not looking at a high spin field. 

Let us for the moment restrict our attention to the GCAs which are obtainable as a limit from relativistic CFTs. 
Our results would be obtainable from the non-relativistic algebra independently as well. Below, we would also present
another method of deriving the results from mode expansions, which in some sense is more intrinsically non-relativistic.
We define the non-relativistic EM tensors in the following way:  
\be{EMtensor}
T_{(1)} = T(z) + \bar {T} (\z) \qquad  T_{(2)} = \e \left[ T(z) - \bar {T} (\z) \right]
\ee  
This definition is a natural choice because of the particular linear combinations of the relativistic Virasoro algebra
which yields the GCA in the non-relativistic limit \refb{Vir2GCA}. Now, as is well known, 
the OPEs among the relativistic EM tensors are of the form
\ben{TT}
&&T(z) T(0) \sim {c/2 \over z^4} + {2 T(0) \over {z^2}} + {\p T(0) \over z} \nonumber\\
&&\bar{T}(\z) \bar{T}(0) \sim {\bar {c}/2 \over \z^4} + {2 T(0) \over {\z^2}} + {\p T(0) \over \z} \\
&&T(z) \bar{T}(0) \sim \mbox{finite} \nonumber  .
\een
which leads to the two-point functions
\be{TT2pt}
\langle T(z) T(0) \rangle = {c \over 2 z^4}, \quad \langle \bar{T}(\z) \bar{T}(0) \rangle = {\bar {c} \over 2 \z^4}, \quad 
\langle T(z) \bar{T}(0) \rangle = 0.
\ee 
To find the non-relativistic answers we use the above results \refb{TT2pt} and perform a calculation similar to 
the limiting calculation of the generic two-point function. We keep in mind $z=t+x, \,\ \z = t-x$ and $c,\bar c = 6 (C_1 \pm C_2/\e)$. 
\begin{eqnarray*}
\langle T_{(1)}(t,x) T_{(1)}(0,0) \rangle = &&\lim_{\e \to 0} 6 t^{-4} \left[(C_1 + {C_2\over \e})  
(1+ 4 {\e x \over t} + \ldots) +  (C_1 - {C_2\over \e}) (1- 4 {\e x \over t} + \ldots )\right] \nonumber\\
\langle T_{(1)}(t,x) T_{(2)}(0,0) \rangle = &&\lim_{\e \to 0} 6 \e t^{-4} \left[(C_1 + {C_2\over \e})  
(1+ 4 {\e x \over t} + \ldots) - (C_1 - {C_2\over \e}) (1- 4 {\e x \over t} + \ldots )\right] \nonumber\\
\langle T_{(2)}(t,x) T_{(2)}(0,0) \rangle = &&\lim_{\e \to 0} 6 \e^2 t^{-4}\left[(C_1 + {C_2\over \e})  
(1+ 4 {\e x \over t} + \ldots) + (C_1 - {C_2\over \e}) (1- 4 {\e x \over t} + \ldots )\right] \nonumber
\end{eqnarray*}
The answers after taking the non-relativistic limit turn out to be 
\ben{nrTT}
\langle T_{(1)}(t,x) T_{(1)}(0,0) \rangle &=& 6 t^{-4} (C_1+ 4 {C_2 x \over t}) \\
\langle T_{(1)}(t,x) T_{(2)}(0,0) \rangle &=&  6 C_2 t^{-4} \\
\langle T_{(2)}(t,x) T_{(2)}(0,0) \rangle &=&  0 
\een
So, here we see that the energy-momentum tensor correlation functions will have non-trivial pieces with spatial dependence 
which survive the scaling limit. Dealing with these would be relatively simpler compared to the arbitrarily high spin states 
in the dual gravity picture. 

Let us re-derive the previous answers in a different way and thereby gain a better understanding of the energy momentum
tensor of the GCA. We turn to mode expansions of the relativistic E-M tensor in terms of the Virasoro algebra generators.
As is well known, we can expand the holomorphic and anti-holomorphic energy-momentum tensors as follows:
\be{mode-Vir}
T(z) = \sum_n \L_n z^{-n-2}, \quad \bar T(\z) = \sum_n \bL_n \z^{-n-2}
\ee
So for the linear combinations which yield the non-relativistic versions of the energy-momentum tensors, 
these expressions become 
\be{T12-mode}
T_{(1)} = \sum_n \L_n z^{-n-2} + \sum_n \bL_n \z^{-n-2}, \quad T_{(2)} = \e (\sum_n \L_n z^{-n-2} - \sum_n \bL_n \z^{-n-2})
\ee
Expanding in terms of $t, x$, we get
\ben{Txt}
T_{(1)} &=& \sum_n t^{-n-2}\bigg( (\L_n + \bL_n) - (n+2) {\e x \over t} (\L_n - \bL_n) + {(n+2)(n+3)\over{2}}({\e x\over t})^2 (\L_n + \bL_n)
  + \ldots \bigg) \nonumber\\
T_{(2)} &=& \e \sum_n t^{-n-2} \bigg( (\L_n - \bL_n) - (n+2) {\e x \over t} (\L_n + \bL_n) + \ldots \bigg) 
\een
Taking the $\e\to 0 $ limit, one gets the mode expansion of the non-relativistic EM tensor in terms of the 
GCA generators.
\be{nrmode}
T_{(1)} = \sum_n t^{-n-2}\bigg[ L_n + (n+2) {x \over t} M_n \bigg], \quad T_{(2)} = \sum_n M_n t^{-n-2}
\ee
Now, we can use \refb{nrmode} to re-calculated the correlation functions of the energy-momentum tensors.
Using \refb{gcawc}, it is straight-forward 
to reproduce \refb{nrTT}. Below, we give the details of one such computation. 
\ben{exp-cal}
&& \hspace{-1cm} \langle T_{(1)} (t_1, x_1) T_{(1)} (t_2, x_2) \rangle = \sum_{m,n} \langle 0|  (L_m + (m+2) {x_1 \over t_1} M_m) t_1^{-m-2} (L_n + 
(n+2) {x_2 \over t_2} M_n) t_2^{-n-2} |0 \rangle \nonumber \\
&=& \sum_{m>1,n<-1} t_1^{-m-2} t_2^{-n-2} \bigg( \langle 0|  L_n L_m + [L_m, L_n]|0 \rangle  
+ (m+2) {x_1\over t_1} \langle 0|  L_n M_m + [M_m, L_n]|0 \rangle \nonumber \\ 
&& \hspace{4cm} + (n+2) {x_2\over t_2} \langle 0|  M_n L_m + [L_m, M_n]|0 \rangle \bigg) \nonumber \\
&=& \sum_{m>1,n<-1} \delta_{m+n,0} (m^3 -m) t_1^{-m-2} t_2^{-n-2} \bigg(C_1 + C_2\{ (m+2) {x_1\over t_1} 
+ (n+2){x_2\over t_2} \} \bigg) \nonumber \\
&=& C_1 \sum_{n=0}^{\infty} \{ (n+2)^3 - (n+2) \} t_1^{-4} ({t_2 \over t_1})^n 
+ C_2 \sum_{n=0}^{\infty} (n+4) \{ (n+2)^3 - (n+2) \} x_1 t_1^{-5} ({t_2 \over t_1})^n \nonumber \\
&& \hspace{2cm} + C_2 \sum_{p=-1}^{\infty} (p+1) \{ (p+3)^3 - (p+3) \} x_2 t_1^{-5} ({t_2 \over t_1})^p \nonumber \\
&=& \sum_{n=0}^{\infty} \bigg( (n+1)(n+2)(n+3) C_1 t_1^{-4} ({t_2 \over t_1})^n 
+ (n+1)(n+2)(n+3)(n+4) C_2 (x_1-x_2) t_1^{-5} ({t_2 \over t_1})^n \bigg) \nonumber \\
& \Rightarrow&\langle T_{(1)} (t_1, x_1) T_{(1)} (t_2, x_2) \rangle = 6 (t_1 -t_2)^{-4} \bigg( C_1 + 4 C_2 {(x_1- x_2) \over (t_1 -t_2) } \bigg)
\een
In \refb{nrTT}, we have used $x_1=x, x_2=0, t_1=t, t_2=0$. Similarly, the rest of the answers of \refb{nrTT} can be reproduced.

\section{The Gravity Dual to the 2D GCA}

In \cite{Bagchi:2009my}, the gravity dual of the GCA was proposed to be a novel Newton-Cartan like $AdS_2 \times R^d$ with
a degenerate metric for the whole of the spacetime and dynamical Christoffel symbols. This was a natural nonrelativistic
extension of the geometrized version of the Newtonian theory in flat spacetimes. For the two dimensional GCA, the dual 
candidate is thus a limit of $AdS_3$ which is a Newton-Cartan like $AdS_2 \times R$. The Brown-Henneaux analysis \cite{BH}
of $AdS_3$ gives rise to two copies of the Virasoro algebra which emerge as the asymptotic symmetries of the $AdS_3$.
The classical GCA (without the central charges) arises on taking the appropriate contraction of these bulk isometries. 
Now, for the quantum GCA with both central charges $C_1, C_2$ turned on, there is more to this story. First, let us 
consider $C_2 =0$. From the Brown-Henneaux analysis in $AdS_3$, central charges for both the Virasoros are generated and
$c_L = c_R = {3l \over 2G}$, where $l, G$ are the radius of $AdS_3$ and the Newton constant in 3d respectively. We see that in
the GCA limit, this is enough to generate $C_1= {l\over 4G}$. As we can see, one would need to look beyond 
the usual $AdS_3$ if one has to realize a non-zero value of $C_2$. But before we venture into this, let us remind
ourselves of a set-up where the usual holographic dictionary works for the non-relativistic theory described by the GCA.

\subsection{Trivial Holography} 

We have seen above that the general two point functions of the non-relativistic field theory depend on time and space 
directions. Let us concentrate for the moment on the sub-sector where we have set all rapidity eigenvalues $\xi = 0$. The correlation functions then
become ultra time dependent and the structure is reminiscent of a one-dimensional CFT. One expects that in this case,
the spatial fibres in the bulk Newton-Cartan structure would not play any role in the holographic description and all 
the information would come from the $AdS_2$ base. This expectation is indeed correct and following \cite{Alishahiha:2009np},
one can consider, e.g. a scalar field in this background whose equation of motion is given by 
\be{scleq}
{1 \over \sqrt{G}}\partial_M\bigg(\sqrt{G}G^{MN}\partial_N\phi(t,z,x_i)\bigg)-m^2\phi(t,z,x_i)=0,
\ee 
where $G_{MN}$ is the metric of $AdS_3$ in the Poincar\'e coordinates. 
The non-relativistic scaling in the bulk is $\{ x \to \e x; t,z \to t,z \}$ where the radial direction 
being a measure of the energy of the boundary theory, scales like time.
Denoting $g_{ab}$ as the metric of $AdS_2$, \refb{scleq} becomes 
\be{diffeq}
\left[{1 \over z^3} \partial_a\bigg(z^3 g^{ab}\partial_b\phi(t,z,x)\bigg)-m^2\phi(t,z,x)\right]
+\frac{z^2}{\epsilon^2}\partial_x^2\phi(t,z,x)=0. 
\ee 
We wish to have a well behaved equation as $\e \to 0$ and hence
\be{well}
{1 \over z^3} \partial_a\bigg( z^3 g^{ab}\partial_b\phi(t,z,x)\bigg)-m^2\phi(t,z,x)=0,\quad
\partial_x^2\phi(t,z,x)=0.
\ee
We could also in principle have a mass term in the $x$-equation if the 3d mass had a $1\over{\e}$ piece. This does not
change the results that follow. The first equation may be obtained from a two dimensional action
given by 
\be{int}
I=\int dt dz
\;\sqrt{G}\;{1 \over 2} \bigg(g^{ab}\partial_a\phi\partial_b\phi+m^2\phi^2\bigg),
\ee 
while the second equation may be treated as a constraint.
As is well known, this can be solved in terms of Bessel functions and the most general solution is 
\be{sol}
\phi(t,z)= z e^{-i\omega t}(A I_\alpha(\omega  z)+B K_\alpha(\omega z)),
\ee 
where $\alpha =\sqrt{m^2 +1} $. Now one can use the usual prescription \cite{Gubser:1998bc, Witten:1998qj} 
of AdS/CFT now for the case of three bulk dimensions and find the bulk solution by given a boundary value
\be{gkpw}
\phi(t,z) = c \delta^{\Delta-2} \int dt'  \phi_\delta(t')
\left( \frac{z}{z^2 +|t-t'|^2} \right)^\Delta,
\ee
where $\Delta=\alpha + {1\over 2}$ and $\phi_\delta$ denotes the
Dirichlet boundary value at $z=\delta$. This can be used to read the
two point function of the boundary GCA: 
\be{f2pt}
\langle \mathcal{O}(t_1) \mathcal{O}(t_2) \rangle \sim
{({t_1-t_2})^{-2\Delta}}.
\ee

\subsection{Topologically Massive Gravity}

We have seen in the previous section how we can apply the usual techniques of holography for the non-relativistic
limit of AdS/CFT for the simple case of zero rapidities ($\xi = 0$). The answers are reminiscent of a truncation 
to an $AdS_2/CFT_1$ correspondence. But for non-zero values of $\xi$, the correlators in the boundary have non-trivial
spatial dependence and to obtain the answers from the bulk, one would need to look at correlation functions of bulk
fields which carry high spins and fixed conformal dimensions. We also want to probe the quantum GCA in all its 
generality and as remarked before, that would mean looking beyond the usual $AdS_3$. It is natural to look for 
corrections to the bulk theory by higher curvature invariants. In three dimensions, there is the unique choice of
adding a gravitational Chern-Simons term to the action. 3d gravity in the absence of any other higher derivative terms
contains no dynamical degrees of freedom. However, adding the Chern-Simons term to the action has the effect of adding 
propagating gravitons to the system and the theory goes under the name of Topologically Massive Gravity (TMG) 
\cite{Deser:1981wh,Deser:1982vy}. The action for TMG is given by
\be{CTMG}
\mathcal{I}=\frac{1}{16\pi G}\int d^3x \sqrt{-g}\left[ R+\frac{2}{\ell^2} + \frac{1}{2\mu} \e^{\mu\nu\rho}
\left(\Gamma^\sigma_{\mu\lambda}\partial_\nu\Gamma^\lambda_{\rho\sigma}+\frac{2}{3}\Gamma^\sigma_{\mu\lambda}
\Gamma^\lambda_{\nu\theta}\Gamma^\theta_{\rho\sigma}\right) \right]
\ee
where $\ell$ is the AdS radius. 
The equations of motion are given by:
\be{eom}
R_{\mu\nu} - {1\over 2} R g_{\mu\nu} + \Lambda g_{\mu\nu} + {1\over \mu} C_{\mu\nu} = 0,
\ee
where $\Lambda = - {1\over \ell^2}$ and the Cotton tensor, $C_{\mu\nu}$ is given by
\be{cotton}
C_{\mu\nu}= \e_{\mu}^{\,\ \alpha \beta} \nabla_{\alpha}( R_{\beta \nu} - {1\over 4} R g_{\beta \nu}). \nonumber
\ee

TMG with its gravitational Chern-Simons term still allows AdS$_3$ as a solution. 
The vacuum state is the globally AdS$_3$
\be{vacuum}
ds^2=-\left(1+\frac{r^2}{\ell^2}\right)dt^2+\left(1+\frac{r^2}{\ell^2}\right)^{-1}dr^2+r^2d\phi^2.
\ee
One also has the BTZ black hole as a solution in TMG. But we would not be interested in it in the present context. 
Since the AdS$_3$ solutions are not changed despite the modification of the equations of motion, 
one can carry out the analogue of the Brown-Henneaux analysis, now for the TMG. One again ends up with two copies of 
the Virasoro algebra \cite{Hotta:2008yq}. 
\ben{globalVir}
\L_n &=& -i e^{inx^+}\left( \p_+ -\frac{n^2\ell^2}{2r^2}\p_- -\frac{inr}{2}\p_r \right), \cr
\bL_n &=& -i e^{inx^-}\left( \p_- -\frac{n^2\ell^2}{2r^2}\p_+ +\frac{inr}{2}\p_r \right)
\een
where $x^\pm=t/\ell\pm \phi$ and $\partial_\pm=\frac{1}{2}(\ell\partial_t\pm \partial_\phi)$.
The topological term however effects the central charges and the effect is an asymmetry between 
the left and right movers \cite{Kraus:2005zm, Kraus:2006wn, Hotta:2008yq}.
\be{cc-tmg}
c =\frac{3\ell}{2G}\left(1+\frac{1}{\mu\ell}\right), \quad \bar c = \frac{3\ell}{2G}\left(1-\frac{1}{\mu\ell}\right)
\ee
Now that we have unequal relativistic central charges, the setting looks ideal to take a nonrelativistic limit \cite{Hotta}. In 
the three dimensional bulk, now in  global co-ordinates, we implement this by taking 
\ben{gravlim}
t\rightarrow t,\quad r\rightarrow r, \quad \phi\rightarrow \epsilon \phi
\quad \textrm{with}\quad
\e\rightarrow0.
\een
It is clear what linear combinations one needs to look at and hence we define the GCA generators $L_n$ and $M_n$ as
\ben{globalGCA}
L_n&=&ie^{int/\ell}\left[\ell\left(1-\frac{n^2\ell^2}{2r^2}\right)\partial_t+in\phi\left(1+\frac{n^2\ell^2}{2r^2}\right)\partial_\phi-inr\partial_r\right],\notag\\
M_n&=&ie^{int/\ell}\left(1+\frac{n^2\ell^2}{2r^2}\right)\partial_\phi.
\een
It is easily to check that the generators above satisfy the center-less GCA algebra.
To obtain both the nonrelativistic central charges, we need to keep in mind that 
the relativistic central charges need to have $1\over \e$ pieces in them which would survive the contraction. 
Hence one must also scale 
\be{mu}
\mu\rightarrow \epsilon \mu.
\ee
Recalling \refb{centch} and \refb{cc-tmg} we get 
\be{centGCA}
C_1 = \frac{\ell}{4G}, \quad C_2 =\frac{1}{4G\mu}.
\ee

\section{TMG/CFT}

We would like to add support to the above relation between the non-relativistic limit of TMG and the GCA, now 
by looking at correlation functions. As stressed earlier, one way of going about this would be to look for 
correlation functions of high spin operators in the bulk. These would have non-trivial spatial dependence. 
To make a connection with TMG, one would need to compute beyond three-point correlators to see the dependence on 
the asymmetric central charges. As suggested before, the way around this rather Herculean task is to look at 
the two-point correlation function of the energy-momentum tensor. Before we go into that, we would like to 
highlight a few points regarding boundary conditions in AdS/CFT which are often glossed over. Our discussion would 
be very close to related discussions in \cite{Skenderis:2009nt, Skenderis:2009kd}.

\subsection{Boundary Conditions and AdS/CFT}
In AdS/CFT, the boundary fields parametrizing the bulk field boundary conditions source the dual operators in
the conformal field theory. The onshell action, through the AdS/CFT correspondence, is the generating functional 
for the CFT correlation functions. One must keep in mind that to obtain the correlation functions of operators in the
dual CFT one must functionally differentiate with respect to the sources that couple to the operators in question. 
These sources should thus be unconstrained and specifically the boundary conditions must not be fixed by hand. 
The leading boundary behaviour should be determined by these unconstrained fields and the equations of motion 
would determine the sub-leading behaviour dynamically. 

For the moment, let us concentrate on the bulk metric as that would be the field of interest for our later discussions.
This should act as the source for the energy momentum tensor of the boundary field theory and hence the boundary
conditions must be parametrized by an unconstrained metric. One needs to consider Asymptotically local AdS (AlAdS)
spacetimes for this purpose. In the finite neighbourhood of the conformal boundary at $r=0$, these spacetimes 
in Gaussian normal co-ordinates centred at the conformal boundary,  admit the following metric
\be{gaussnorm}
ds^2 = {dr^2 \over r^2} + {1 \over r^2} g_{ij}(x,r) dx^i dx^j,
\ee
where $g_{ij}(x,r) \to g_{(0)ij}(x)$ as $r \to 0$. ($g_{(0)ij}(x)$ is a non-degenerate metric). The precise form of
$g_{ij}(x,r)$ is determined by solving the bulk equations asymptotically and for pure gravity in 3d, this yields
\be{pure3d}
g_{ij}(x,r) = g_{(0)ij}(x) + r^2 g_{(2)ij}(x) + \ldots 
\ee
Now, the analysis of Brown and Henneaux \cite{BH} imposes the further criterion on the boundary conditions:
\be{BH3}
g_{(0)ij}(x) = \delta_{ij}
\ee
that implies that the metric should be asymptotically AdS. This, from the point of AdS/CFT is not general enough and 
is violated if one wishes to consider CFTs in non-trivial backgrounds or wishes, as we do, to calculate correlation 
functions of the stress-energy tensor. 

\subsection{Holographic Renormalization and Correlation functions}
Correlation functions in the field theory suffer from ultraviolet (UV) divergences. These are expected to be 
related to the infrared (IR) divergences on the gravity side by the UV-IR correspondence in AdS/CFT. To get 
finite, sensible answers in the field theory, one needs to regularize and renormalize. Similarly, 
one would need to ``holographically renormalize'' in the bulk \cite{de Haro:2000xn, Skenderis:2002wp}. 
The calculation outlined for the bulk dual of GCA, for scalar operators in the regime where $\xi=0$ in Sec (4.1) 
is an example of implementation of the GKPW prescription \cite{Gubser:1998bc, Witten:1998qj} in the bulk which
should ideally be looked at as a relation between bare quantities. In principle, to get proper answers one should 
use the techniques of holographic renomalization. Holographic renomalization of the onshell gravity action 
depends crucially on the asymptotic form of the metric and hence the boundary conditions talked about earlier become
important in this procedure. To holographically renormalize, one adds local boundary covariant counter-terms.
For AlAdS spacetimes, the procedure yields the one-point function of the energy momentum tensor 
\be{1pt}
\langle T_{ij} \rangle \sim g_{(d)ij} + X_{ij} [g_{(0)}] 
\ee
where $X_{ij} [g_{(0)}]$ is a known function of $g_{(0)}$. One finds the higher point correlation functions of 
the energy-momentum tensor by taking derivatives of the above relation. 

\subsection{Holographic Renormalization and TMG}
An analogous discussion to the above holds for the TMG and one can show that the Brown-Henneaux boundary conditions
are not sufficiently general. In particular, at the ``chiral'' point $\mu \ell =1$, where one of the central charges of the 
relativistic theory (see \refb{cc-tmg}) vanishes, there is the existence of extra logarithmic modes \cite{Grumiller:2008qz}. This invalidates the arguments of 
the chiral gravity conjecture put forward in \cite{Li:2008dq}. We shall not dwell on these aspects in this paper. 

We are looking to calculate correlation functions of the energy momentum tensor. For an honest calculation, one 
needs to adopt the techniques of holographic renormalization to this context. This has fortunately been dealt with 
in \cite{Skenderis:2009nt} in great details. Let us here point out a few new features on this set-up compared to the 
usual scenario in $AdS$ spacetimes.  
The field equations for the TMG are third order in derivatives. This one can fix two pieces of boundary data: 
the metric and a component of the extrinsic curvature. The boundary metric sources the EM tensor while the 
boundary field parametrizing the extrinsic curvature sources a new operator. This new operator turns out to be 
the logarithmic partner of the energy momentum tensor in the chiral limit. The dual of the TMG at $\mu =1$ is 
thus a logarithmic CFT with $c=0$. We would however not have much to say about the role of this operator in the 
non-relativistic limit and will entirely focus on the energy momentum tensor. 

\subsection{TMG and GCA}

For a general $\mu$, the authors \cite{Skenderis:2009nt} compute the two point function of the energy-momentum tensor
by using the techniques of holographic renormalization. We shall not go into the details of the derivation here and 
quote the results and proceed to use them in our context. The authors \cite{Skenderis:2009nt} use a linearized analysis
and expand $g_{ij} = \eta_{ij} + h_{ij}$. They work in Poincare co-ordinates where the background metric is 
\be{backgr}
G_{\mu \nu} dx^{\mu} dx^{\nu} = {dr^2 \over r^2} + {1 \over r^2} \eta_{ij} dx^i dx^j = {dr^2 \over r^2} + {1 \over r^2} dz d\z
\ee
where one first uses lightcone co-ordinates $v,u = t \pm x$ and then replaces $(v,u) \to (z, \z)$, the complex 
boundary co-ordinates. 
The results for the two-point function of the energy-momentum tensor for a general $\mu$ is 
\ben{hol-ren-TT}
\langle \bar T(z, \z) \bar T(0) \rangle &=& {3\over 2G_N}{\mu + 1 \over 2 \mu} {1\over \z^4} \crcr
\langle T(z, \z) T(0) \rangle &=& {3\over 2G_N}{\mu - 1 \over 2 \mu} {1\over z^4}
\een
To realize the non-relativistic results, we take linear combinations as before. This yields
\ben{TT1}
\langle T_{(1)}(t,x)  T_{(1)}(0)\rangle &=& \langle [T(z, \z) + \bar T(z, \z)] [T(0) + \bar T(0)] \rangle \crcr 
= {3\over 2G_N} t^{-4} \bigg[\hspace{-0.5cm}&&(1 + 10 ({x \over t})^2 + 35 ({x \over t})^4 + \ldots ) + {1\over \mu} (4 ({x \over t}) 
+ 20 ({x \over t})^3 + \ldots) \bigg]
\een
Similarly, 
\ben{TT2}
\langle T_{(1)}(t,x)  T_{(2)}(0)\rangle &=& \e \langle [T(z, \z) + \bar T(z, \z)] [T(0) - \bar T(0)] \rangle \crcr 
= {3\e\over 2G_N} t^{-4} \bigg[\hspace{-0.5cm}&& (4 ({x \over t}) + 20 ({x \over t})^3 + \ldots) 
+ {1\over \mu} (1 + 10 ({x \over t})^2 + 35 ({x \over t})^4 + \ldots )\bigg] 
\een
And,
\ben{TT12}
\langle T_{(2)}(t,x)  T_{(2)}(0)\rangle &=& \e^2 \langle [T(z, \z) - \bar T(z, \z)] [T(0) - \bar T(0)] \crcr 
= {3\e^2 \over 2G_N} t^{-4} \bigg[\hspace{-0.5cm}&&(1 + 10 ({x \over t})^2 + 35 ({x \over t})^4 + \ldots ) + {1\over \mu} (4 ({x \over t}) 
+ 20 ({x \over t})^3 + \ldots) \bigg]
\een

When we take the non-relativistic limit of $x\to \e x, t\to t$ and $\e \to 0$, we see that there is no non-trivial 
spatial dependence in \refb{TT1} if we do not scale $\mu$. To match \refb{TT1} -- \refb{TT12} to the calculations 
performed purely in the non-relativistic boundary theory \refb{nrTT}, we see that the coefficient of the gravitational
Chern-Simons term $\mu$ must scale like $\mu \to \e \mu$ as $\e \to 0$, as proposed in \refb{mu} \cite{Hotta}. 
The calculation in \cite{Skenderis:2009nt} was performed in the range $0<|\mu| < 2$. Later, these techniques have 
been extended to $\mu = 3$ which is of relevance to Schrodinger holography \cite{Guica:2010sw}. But there is no obstruction
in extending these results to the limit where $\mu$ goes to zero. \footnote{When $\mu$ takes integral
values, in our case the strict $\mu =0$ there are additional divergences, and additional contributions to 
one point functions, which are local in the sources. These terms contribute only contact terms to the 
two point functions. The identification of the central charges below continues to hold. We thank Marika Taylor
for helpful correspondence on this issue.}
\ben{TTgrav}
\langle T_{(1)}(t,x)  T_{(1)}(0)\rangle &=& {3\over 2G_N} t^{-4} \bigg[1 + {4\over \mu}  ({x \over t}) \bigg] \crcr
\langle T_{(1)}(t,x)  T_{(2)}(0)\rangle &=& {3\over 2G_N \mu} t^{-4} \crcr
\langle T_{(2)}(t,x)  T_{(2)}(0)\rangle &=& 0.
\een

The identifications also lead to central terms of the 2d quantum GCA in terms of the bulk parameters as:
\be{gcacc}
C_1 = {1\over 4G} , \quad C_2 = {1\over 4G \mu}
\ee
in agreement with the proposal of \cite{Hotta}.   

\section{Conclusions and future directions}

In this paper, we have reviewed our understanding of aspects of Galilean holography where 
the Galilean Conformal Algebra is the symmetry algebra of the boundary theory. We have specifically looked at 
the 2d GCA and pointed out novel spatial dependence that exists in the correlation functions of the 
non-relativistic energy momentum tensor. This is an instance where one does not need to consider 
a setting where the parent relativistic operators have arbitrarily high spins. The construction of the bulk calculation
becomes much simpler as one does not have to look at correlation functions of arbitrary high spin fields in AdS.    
In the bulk calculation, we have stressed why we need to look beyond a Brown-Henneaux like boundary conditions when
we are computing stress-energy correlators and why the framework of holographic renormalization is the correct set-up
to use. We discussed why we need to go beyond $AdS_3$ and look at Topologically Massive 3d Gravity if we wanted to 
realize the 2d GCA in its full generality. Finally, we relied heavily on the techniques and results of \cite{Skenderis:2009nt} to show that in the non-relativistic
limit, to obtain the non-trivial spatial dependence obtained in calculation in the nonrelativistic boundary theory. 
We found that one needs to scale the co-efficient of the gravitational Chern-Simons term, as proposed independently 
in \cite{Hotta}. Our analysis expectedly also yields the corrected central charges.  

This is in some sense a first step towards understanding TMG/GCA holography. The process of holographic renormalization
has been outlined in all generality for the TMG in \cite{Skenderis:2009nt}. It should, in principle, be possible
to take the nonrelativistic limit at any stage and follow the steps of the argument to see if the procedure makes 
sense. We know for certain now that taking the limit at the end yields perfectly sensible answers which match with 
calculations purely in terms of the nonrelativistic boundary theory. It is plausible that following the results of 
limits taken at various points in the calculation, we would be able to learn more about the structure of the bulk 
theory. This is under current investigation. 

There is also the rather intriguing aspect of the ``chiral point'' $\mu \ell = 1$ where one of the central charges
of the relativistic CFT becomes zero. This has been shown to be a logarithmic CFT. From the point of view of 
the non-relativistic set-up, the chiral point could be viewed as a ``double scaling'' limit, where one takes 
$\mu \to \e \mu$, $\ell \to {1\over \e} \ell$ keeping $\mu \ell = 1$. One knows that the aforementioned techniques of 
holographic renormalization break down for asymptotically flat spacetimes. But it is possible one might get away by 
keeping a very small cosmological constant in this case because we {\emph{do not}} need to put $\mu =0$ strictly anywhere. 
The result should be a logarithmic version of the GCA. We have not looked at the operator in the CFT which is sourced 
by the extrinsic curvature which one needs to turn on in TMG. This, as pointed out earlier, plays the role of the 
log-partner of the energy momentum tensor in the usual relativistic case. It is clear that this would also have a vital
role to play in the non-relativistic set-up. It is also curious to note that the $\ell \to {1\over \e} \ell$ limit of 
symmetries of $AdS_3$ leads to the BMS algebra which is isomorphic to the 2d GCA \cite{Bagchi:2010}. It would be 
interesting to understand how the whole story fits. 

To conclude, let us point out something that we have glossed over in our discussions so far. It is well known that 
Topologically Massive Gravity is plagued with aspects of non-unitarity, coming from negative energy of the propagating
gravitons. In the limit of large $\mu$ and the non-relativistic context, this problem persists and is reflected in 
the field theory in the generic non-unitarity of the GCA \cite{Bagchi:2009pe}. We should point out that since we
are dealing with finite relativistic spins to start off with, in the non-relativistic sector we are in the $\xi=0$ 
sector. But we are also in the $C_2 \ne 0$ sector. There exists null vectors in this GCA sector and interestingly, the 
operator $M_0$ has a block diagonal form reminiscent of logarithmic CFTs \cite{Bagchi:2009pe}. It would be of interest
to explore the behaviour of the operator $X_{ij}$ which sources the extrinsic curvature on the gravity side even from this
perspective and understand the nature of the non-unitarity in the non-relativistic context.

\acknowledgments{It is a pleasure to thank Rajesh Gopakumar, Rajesh Gupta, Patricia Ritter, Joan Simon and Marika Taylor for 
helpful discussions and correspondence. The author would also like to thank the Institute for Advanced Studies, Princeton
for the warm hospitality extended to him during the initial part of this work.}


\begin{thebibliography}{999}

\bibitem{Maldacena:1997re}
  J.~M.~Maldacena,
  ``The large N limit of superconformal field theories and supergravity,''
  Adv.\ Theor.\ Math.\ Phys.\  {\bf 2}, 231 (1998)
  [Int.\ J.\ Theor.\ Phys.\  {\bf 38}, 1113 (1999)]
  [arXiv:hep-th/9711200].

\bibitem{Hagen:1972pd}
  C.~R.~Hagen,
  ``Scale and conformal transformations in galilean-covariant field theory,''
  Phys.\ Rev.\  D {\bf 5}, 377 (1972).

\bibitem{Niederer:1972zz}
  U.~Niederer,
  ``The maximal kinematical invariance group of the free Schrodinger
  equation,''
  Helv.\ Phys.\ Acta {\bf 45}, 802 (1972).


\bibitem{Henkel:1993sg}
  M.~Henkel,
  ``Schr\"odinger invariance in strongly anisotropic critical systems,''
  J.\ Statist.\ Phys.\  {\bf 75}, 1023 (1994)
  [arXiv:hep-th/9310081].

\bibitem{Nishida:2007pj}
  Y.~Nishida and D.~T.~Son,
  ``Nonrelativistic conformal field theories,''
  Phys.\ Rev.\  D {\bf 76}, 086004 (2007)
  [arXiv:0706.3746 [hep-th]].


\bibitem{Son:2008ye}
  D.~T.~Son,
  ``Toward an AdS/cold atoms correspondence: a geometric realization of the
  Schroedinger symmetry,''
  Phys.\ Rev.\  D {\bf 78}, 046003 (2008)
  [arXiv:0804.3972 [hep-th]].

\bibitem{Balasubramanian:2008dm}
  K.~Balasubramanian and J.~McGreevy,
  ``Gravity duals for non-relativistic CFTs,''
  Phys.\ Rev.\ Lett.\  {\bf 101}, 061601 (2008)
  [arXiv:0804.4053 [hep-th]].
  
\bibitem{Hartnoll}
  S.~A.~Hartnoll,
  ``Lectures on holographic methods for condensed matter physics,''
  Class.\ Quant.\ Grav.\  {\bf 26}, 224002 (2009)
  [arXiv:0903.3246 [hep-th]].
  
\bibitem{negro1997}
J.~Negro, M.~A.~del Olmo, and A.~Rodr�guez-Marco.
"Non-relativistic Conformal Groups I,"
J.\ Math.\ Phys {\bf 38}, 3786 (1997).


\bibitem{Lukierski:2005xy}
  J.~Lukierski, P.~C.~Stichel and W.~J.~Zakrzewski,
  ``Exotic Galilean conformal symmetry and its dynamical realisations,''
  Phys.\ Lett.\  A {\bf 357}, 1 (2006)
  [arXiv:hep-th/0511259].

\bibitem{Gomis:2005pg}
  J.~Gomis, J.~Gomis and K.~Kamimura,
  ``Non-relativistic superstrings: A new soluble sector of AdS(5) x S**5,''
   JHEP {\bf 0512}, 024 (2005)
  [arXiv:hep-th/0507036].
  
\bibitem{Bagchi:2009my}
  A.~Bagchi and R.~Gopakumar,
  ``Galilean Conformal Algebras and AdS/CFT,''
  JHEP {\bf 0907}, 037 (2009)
  [arXiv:0902.1385 [hep-th]].

\bibitem{Duval:2009vt}
  C.~Duval and P.~A.~Horvathy,
  ``Non-relativistic conformal symmetries and Newton-Cartan structures,''
  J.\ Phys.\ A  {\bf 42}, 465206 (2009)
  [arXiv:0904.0531 [math-ph]].

\bibitem{Alishahiha:2009np}
  M.~Alishahiha, A.~Davody and A.~Vahedi,
  ``On AdS/CFT of Galilean Conformal Field Theories,''
  arXiv:0903.3953 [hep-th].

\bibitem{Bagchi:2009ca}
  A.~Bagchi and I.~Mandal,
  ``On Representations and Correlation Functions of Galilean Conformal
  Algebras,''
  Phys.\ Lett.\  B {\bf 675}, 393 (2009)
  [arXiv:0903.4524 [hep-th]].

\bibitem{Bagchi:2009pe}
  A.~Bagchi, R.~Gopakumar, I.~Mandal {\it et al.},
  ``GCA in 2d,''
  JHEP {\bf 1008}, 004 (2010).
  [arXiv:0912.1090 [hep-th]].

\bibitem{IPM}
  A.~Hosseiny and S.~Rouhani,
  ``Affine Extension of Galilean Conformal Algebra in 2+1 Dimensions,''
  arXiv:0909.1203 [hep-th].

\bibitem{Hotta}
  K.~Hotta, T.~Kubota and T.~Nishinaka,
  ``Galilean Conformal Algebra in Two Dimensions and Cosmological Topologically
  Massive Gravity,''
  arXiv:1003.1203 [hep-th].


\bibitem{Bagchi:2010}
  A.~Bagchi,
  ``Correspondence between Asymptotically Flat Spacetimes and Nonrelativistic Conformal Field Theories,''
  Phys.\ Rev.\ Lett.\  {\bf 105}, 171601 (2010).

  A.~Bagchi,
  ``The BMS/GCA correspondence,''
  [arXiv:1006.3354 [hep-th]].

\bibitem{Bagchi:2010xw}
  A.~Bagchi and A.~Kundu,
  ``Metrics with Galilean Conformal Isometry,''
  arXiv:1011.4999 [hep-th].

\bibitem{BH}
  J.~D.~Brown, M.~Henneaux,
  ``Central Charges in the Canonical Realization of Asymptotic Symmetries: An Example from Three-Dimensional Gravity,''
  Commun.\ Math.\ Phys.\  {\bf 104}, 207-226 (1986).

\bibitem{Deser:1982vy}
  S.~Deser, R.~Jackiw, S.~Templeton,
  ``Three-Dimensional Massive Gauge Theories,''
  Phys.\ Rev.\ Lett.\  {\bf 48}, 975-978 (1982).
  
\bibitem{Deser:1981wh}
  S.~Deser, R.~Jackiw, S.~Templeton,
  ``Topologically Massive Gauge Theories,''
  Annals Phys.\  {\bf 140}, 372-411 (1982).

\bibitem{Kraus:2005zm}
  P.~Kraus, F.~Larsen,
  ``Holographic gravitational anomalies,''
  JHEP {\bf 0601}, 022 (2006).
  [hep-th/0508218].

\bibitem{Kraus:2006wn}
  P.~Kraus,
  ``Lectures on black holes and the AdS(3) / CFT(2) correspondence,''
  Lect.\ Notes Phys.\  {\bf 755}, 193-247 (2008).
  [hep-th/0609074].

\bibitem{Hotta:2008yq}
  K.~Hotta, Y.~Hyakutake, T.~Kubota {\it et al.},
  ``Brown-Henneaux's Canonical Approach to Topologically Massive Gravity,''
  JHEP {\bf 0807}, 066 (2008).
  [arXiv:0805.2005 [hep-th]].

\bibitem{Skenderis:2009nt}
  K.~Skenderis, M.~Taylor and B.~C.~van Rees,
  ``Topologically Massive Gravity and the AdS/CFT Correspondence,''
  JHEP {\bf 0909}, 045 (2009)
  [arXiv:0906.4926 [hep-th]].

\bibitem{Skenderis:2009kd}
  K.~Skenderis, M.~Taylor, B.~C.~van Rees,
  ``AdS boundary conditions and the Topologically Massive Gravity/CFT correspondence,''
  [arXiv:0909.5617 [hep-th]].

\bibitem{Guica:2010sw}
  M.~Guica, K.~Skenderis, M.~Taylor {\it et al.},
  ``Holography for Schrodinger backgrounds,''
  [arXiv:1008.1991 [hep-th]].

\bibitem{Gubser:1998bc}
  S.~S.~Gubser, I.~R.~Klebanov, A.~M.~Polyakov,
  ``Gauge theory correlators from noncritical string theory,''
  Phys.\ Lett.\  {\bf B428}, 105-114 (1998).
  [hep-th/9802109].

\bibitem{Witten:1998qj}
  E.~Witten,
  ``Anti-de Sitter space and holography,''
  Adv.\ Theor.\ Math.\ Phys.\  {\bf 2}, 253-291 (1998).
  [hep-th/9802150].


\bibitem{de Haro:2000xn}
  S.~de Haro, S.~N.~Solodukhin, K.~Skenderis,
  ``Holographic reconstruction of space-time and renormalization in the AdS / CFT correspondence,''
  Commun.\ Math.\ Phys.\  {\bf 217}, 595-622 (2001).
  [hep-th/0002230].

\bibitem{Skenderis:2002wp}
  K.~Skenderis,
  ``Lecture notes on holographic renormalization,''
  Class.\ Quant.\ Grav.\  {\bf 19}, 5849-5876 (2002).
  [hep-th/0209067].

\bibitem{Grumiller:2008qz}
  D.~Grumiller, N.~Johansson,
  ``Instability in cosmological topologically massive gravity at the chiral point,''
  JHEP {\bf 0807}, 134 (2008).
  [arXiv:0805.2610 [hep-th]].

\bibitem{Li:2008dq}
  W.~Li, W.~Song, A.~Strominger,
  ``Chiral Gravity in Three Dimensions,''
  JHEP {\bf 0804}, 082 (2008).
  [arXiv:0801.4566 [hep-th]].





\end{thebibliography}
\end{document}